\documentclass[conference]{IEEEtran}
\IEEEoverridecommandlockouts
\usepackage{cite}
\usepackage{amsmath,amssymb,amsfonts}
\usepackage{algorithmic}
\usepackage{graphicx}
\usepackage{textcomp}
\usepackage{xcolor}
\usepackage{footnote}
\usepackage{multirow}
\usepackage{booktabs}
\usepackage{caption}
\usepackage{subcaption}
\usepackage{csquotes}
\usepackage[inline]{enumitem}
\usepackage{url}
\usepackage{comment}
\usepackage{listings}
\usepackage{footnote}
\usepackage{url}
\usepackage{float}
\usepackage{marvosym}
\usepackage{pifont}
\usepackage{xcolor}
\usepackage{colortbl}
\usepackage{balance}

\usepackage{todonotes}

\definecolor{gray(x11gray)}{rgb}{0.75, 0.75, 0.75}
\newcommand{\tieGray}{\cellcolor{gray}}

\newcommand{\todoi}[1]{\todo[inline]{\small #1}}

\newcommand{\sfreqEfNew}{$\phi$}

\newcommand{\conf}{$\mathit{Confidence}$}
\newcommand{\sconf}{$\mathit{Conf}$}
\newcommand{\dstar}{$\mathit{DStar}$}
\newcommand{\gpA}{$\mathit{GP13}$}

\newcommand{\ochiai}{$\mathit{Ochiai}$}
\newcommand{\tarantula}{$\mathit{Tarantula}$}

\newcommand{\sdstar}{$\mathit{DSt}$}

\newcommand{\sochiai}{$\mathit{Och}$}
\newcommand{\starantula}{$\mathit{Tar}$}

\newcommand{\cmark}{\textcolor{green!80!black}{\ding{51}}}
\newcommand{\xmark}{\textcolor{red}{\ding{55}}}

\def\BibTeX{{\rm B\kern-.05em{\sc i\kern-.025em b}\kern-.08em
    T\kern-.1667em\lower.7ex\hbox{E}\kern-.125emX}}

\newif\ifanonym

\anonymfalse

\begin{document}

\title{Method Calls Frequency-Based Tie-Breaking Strategy For Software Fault Localization}

\ifanonym

\author{\IEEEauthorblockN{Anonymous}}

\else

\author
{

\IEEEauthorblockN{Qusay Idrees Sarhan\textsuperscript{1,2},
B\'ela Vancsics\textsuperscript{1}, \'Arp\'ad Besz\'edes\textsuperscript{1}}
\IEEEauthorblockA{\textsuperscript{1} Department of Software Engineering, University of Szeged, Szeged, Hungary}
\IEEEauthorblockA{\textsuperscript{2} Department of Computer Science, University of Duhok, Duhok, Iraq\\
\{sarhan, vancsics, beszedes\}@inf.u-szeged.hu}

}

\fi 

\maketitle

\begin{abstract}
In Spectrum-Based Fault Localization (SBFL), a suspiciousness score is assigned to each code element based on test coverage and test outcomes.
The scores are then used to rank the code elements relative to each other in order to aid the programmer during the debugging process when seeking the source of a fault.
However, probably none of the known SBFL formulae are guaranteed to produce different scores for all the program elements, hence {\em ties} emerge between the code elements.
Based on our experiments, ties in SBFL are prevalent: in Defects4J, 54--56\% of buggy methods are members of ties, i.e., there is at least one other method with the same score in these cases (but typically much more, on average 6), and this inevitably reduces the effectiveness of any SBFL approach.
In this work, we present a technique to break ties in such cases based on the so-called {\em method calls frequencies}.
This counts the number of different contexts of method calls (both as callees and as callers) in failing test cases.
The intuition is that if a method appears in many different calling contexts during a failing test case, it will be more suspicious and get a higher rank position compared to other methods with the same scores.
This method can be applied to any underlying SBFL formula, and can favourably break the occurring ranks in the ties in many cases. The experimental results show that our novel tie-breaking strategy achieved a significant reduction in both size and number of critical ties in our benchmark. In 72-73\% of the cases, the ties were completely eliminated and the average reduction rate was more than 80\%. 
\end{abstract}


\section{Introduction}

During debugging, fault localization is one of the most difficult and time-consuming tasks, particularly for large-scale software systems. Therefore, there is a high demand for automatic fault localization techniques that can help software engineers effectively find the locations of faults with minimal human intervention ~\cite{intro1}. This has led to propose and implement different types of such techniques. Spectrum Based Fault Localization (SBFL) is considered amongst the most prominent techniques in this respect due to its efficiency and effectiveness ~\cite{intro2}. In SBFL, the probability of each program entity (e.g., statements, blocks, or methods) of being faulty is calculated based on test cases, their results, and their corresponding code coverage information. 

Unfortunately, SBFL techniques are not yet widely adopted in the industry ~\cite{intro3, intro4, survey3} because they pose a number of issues and their performance is affected by several influencing factors. One of these factors is the following. In SBFL, program statements are ranked in order of their suspiciousness from most suspicious to least. To decide whether a statement is faulty or not, programmers examine each statement starting from the top of the ranking. In order to help developers discover the faulty statement early in the examination process and with minimal effort, the faulty statement should be put near to the highest place in the ranking. However, ranking based only on suspiciousness scores inevitably involves a problem called {\em rank ties}~\cite{ref2}. When different code elements (such as statements or methods) are tied this means that they have the same suspiciousness scores, so they are indistinguishable from each other in this respect.
If the fauly element falls within a tie (this is called a {\em critical tie}) then the overall performance of the SBFL method will be reduced.

Probably none of the known SBFL formulae are guaranteed to produce different scores for all the program elements, hence  ties inevitably emerge between the code elements.
In fact, as we shall see in this paper, ties in SBFL are prevalent regardless of the underlying formula. 
In this paper, we propose a tie-breaking strategy to improve the performance of SBFL by utilizing contextual information extracted from method call chains (our strategy is at method-level granularity, meaning that the basic program element considered for fault localization is a method). Method call chains are the call sequences of methods in the call stack during their executions. Both call chains and call stack traces can provide valuable context to the fault being traced. For example, a method may fail if called from one place and performs successfully when called from another.

The proposed strategy is based on how often a method has been called, directly or indirectly, during the execution of failed test cases {\em in different contexts}. However, here we do not count all occurrences of a method call but only those that occur in unique call contexts. Thus, repeating sequences of method calls due to, e.g., loops are not considered.  
The intuition is that if a method is present in many different calling contexts during a failing test case, it will be more suspicious and get a higher rank position compared to other methods with the same scores.
The strategy can be applied to any underlying SBFL formula, and, as we will see, it can favourably break the occurring ranks in the ties in many cases.

We empirically evaluated the approach using 411 real faults from the Defects4J dataset and five well-known spectra formulae. 
The obtained results indicate that for all the selected formulae, the call frequency chain based tie-breaking strategy can improve the localization effectiveness in many ways. For example, it completely eliminated 72--73\% of the critical ties over the full dataset. In other cases, it reduced their sizes significantly.
Ranks of buggy elements improved by two positions on average, the approach achieved positive movement of bug ranks in most Top-3/Top-5/Top-10 rank categories, and in particular, the number of cases where the faulty method became the top ranked element increased by 23--30\%.

The main contributions in the paper can be summarized as follows:
\begin{enumerate}
\item Analysis of rank tie prevalence in the benchmark programs.
\item A new tie-breaking algorithm that successfully breaks critical ties in many cases.
\item The analysis of the impact of tie-breaking on the overall SBFL effectiveness.
\end{enumerate}

In terms of the concrete research goals, we defined the following Research Questions (RQs) for this paper:
\begin{itemize}
\item[\bf RQ1] How prevalent are rank ties when applying a selection of different SBFL formulae? In particular: \\
How common are rank ties in the Defects4J benchmark and what are their sizes? \\
What would be the theoretically achievable maximum improvement if all critical ties were broken?
\item[\bf RQ2] What level of tie-breaking can we achieve using the call-frequency based strategy?
\item[\bf RQ3] What is the overall effect of the proposed tie-breaking on SBFL effectiveness in terms of global rank improvement?
\end{itemize}

The remainder of the paper is organized as follows. Section~\ref{related_work} presents an overview of the related work. Section~\ref{fl_ties} describes the tie problem in software fault localization. Section~\ref{tie_breaking} deals with RQ1, while Section~\ref{freq_ef} introduces our novel tie-breaking approach and answers RQ2. Section~\ref{results} presents the description of our empirical evaluation of RQ3. Section~\ref{threats} reports the potential threats to validity, finally we provide our conclusions in Section~\ref{conclusions}.

\section{Related Work}
\label{related_work}

Software fault localization is a significant research topic in software engineering. Despite having started in the late 1950s, software fault localization research has gained more attention in the last couple decades. This is reflected in the increase in the number of techniques, tools, and publications. The main reason for the increased attention is the dramatic increase in software systems size due to the newly added functionalities and features they provide. This also has led to an increase in the complexity of these systems. As a result, more faults have also been reported. Here, software fault localization is a good approach to reduce the number of faults and to ensure software quality. Many fault localization techniques, in addition to the ones used in this paper, have been proposed and discussed in the literature. There have been several surveys
written \cite{survey1, survey2, survey3} and various empirical studies \cite{empir1, empir2} performed to compare the effectiveness of various techniques. However, a systematic research work on the problem of addressing ties in the context of fault localization is still modest. 
The most related publications are presented here.

Yu et al.~\cite{ref7} proposed a tie-breaking strategy that firstly sorts program statements based on their suspiciousness and then breaks ties by sorting statements based on applying a confidence metric. The metric is intended to assess the degree of certainty in a given suspiciousness value. For example, when two or more statements are assigned the same level of suspicion, the suspiciousness assigned to the statements with a higher level of certainty is more reliable. As a result, the corresponding statements are more likely to be faulty.

Xu et al.~\cite{ref4} have presented the most systematic analysis of the problem associated with critical ties (ties with faulty statements) where four tie-breaking strategies were considered and evaluated via experimental case studies. Their results indicated that some of the strategies can reduce ties without having an adverse impact on fault localization effectiveness. Besides, they proposed some other tie-breaking techniques to be studied and evaluated in the future such as slicing-based approach to breaking ties.

Debroy et al.~\cite{ref2} proposed a grouping-based strategy that employs another influential factor alongside statements’ suspiciousness. This strategy groups program statements based on the number of failed tests that execute each statement and then sorts the groups that contain statements that have been executed by more failed tests. Afterwards, it ranks the statements within each group by their suspiciousness to generate the final ranking list. Thus, the statements are examined firstly based on their group order and secondly based on their suspiciousness. Their results show that ranking based on several factors can improve the SBFL effectiveness. Thus, the grouping-based strategy could be effective in tie-breaking as well.

Laghari et al.~\cite{method_calls} employed the idea of utilizing method calls to improve the performance of SBFL. In their proposed approach, they combined method calls and their sequences with program slicing to extract spectra patterns from different contexts that can be used to effectively locate faults compared to only using the standard SBFL formulae.

It can be noted that utilizing method calls to improve the performance of SBFL is not new. However, using method calls frequency for tie-breaking is a novel approach which has not been investigated by other researchers previously.


\section{Spectrum-Based Fault Localization and Ties}
\label{fl_ties}

In this section, we present how SBFL techniques are used to locate faults by ranking program elements based on their suspiciousness of being faulty and what are the steps to do so. 
Also, we introduce the problem of ties among program elements in the rankings that these techniques produce. 
This is achieved by a simple code example that illustrate the aforementioned concepts.

\subsection{Fault Localization Formulae}
SBFL is a dynamic program analysis technique which is performed through program execution.
In SBFL, code coverage information (also called spectra) obtained from executing a set of test cases and test results are used to calculate the probability of each program entity (e.g., statements, blocks, or methods) of being faulty \cite{ref15}.
Code coverage provides information on which program entity has been executed and which one has not during the execution of each test case; while tests results are classified as passed or failed test cases.
Passed test cases are executions of a program that output as expected, whereas failed test cases are executions of a program that output as unexpected \cite{ref6}.

\begin{figure}[H]
\caption{Running example -- program code and test cases}
	\label{chain-excode}
	\centering
	\includegraphics[width=8cm, height=10.5cm]{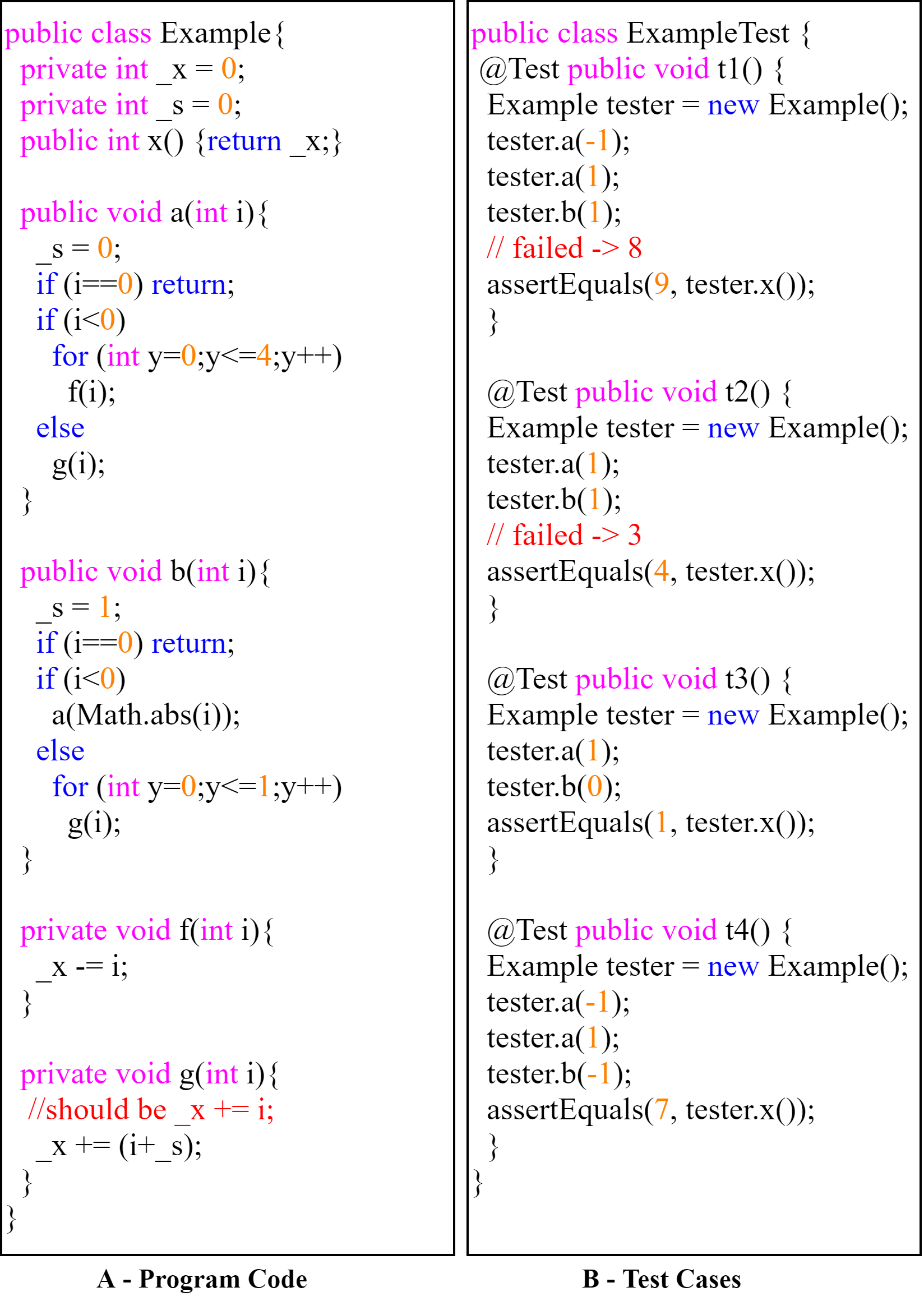}
		\vspace*{-4mm}
\end{figure}

To illustrate the work of SBFL, assume a simple Java program, which is adopted from Vancsics et al.~\cite{VHS21}, that comprises of four main methods ($a$, $b$, $f$, and $g$), and its four test cases ($t1$, $t2$, $t3$, and $t4$) as shown in Figure~\ref{chain-excode}. 

Suppose that the tests have been executed on the program and the program spectra (the execution information of the four program methods in passed and failed test cases) have been recorded. Table~\ref{cov_matrix_and_stat} presents this information. An entry of 1 in the cell corresponding to the method $a$ and the test case $t1$ means that the method $a$ has been executed by the test case $t1$, and 0 otherwise.
This is also known as the {\em hit-based} SBFL.
An entry of 1 in the row labeled ``results'' means that the corresponding test case resulted in failure, and 0 otherwise. For example: $t2$ test case calls the methods $a$, $b$ and $g$ and it failed because its expected value is $3$ and not $4$.

The program spectra are then used by a spectra formula to compute the suspiciousness of each program element of being faulty. Often, a spectra formula is expressed in terms of four counters that are calculated from the program spectra as follows:

\begin{enumerate}[label=\emph{\alph*)}]
	\item \textbf{$m_{ef}$}: set of failed test cases that executed \textit{m}.
	\item \textbf{$m_{ep}$}: set of passed test cases that executed \textit{m}.
	\item \textbf{$m_{nf}$}: set of failed test cases that not executed \textit{m}.
	\item \textbf{$m_{np}$}: set of passed test cases that not executed \textit{m}.\\
\end{enumerate}

The last four columns of Table~\ref{cov_matrix_and_stat} represent these values. For example, $\mathit{ef}$ of $a$ contains two tests because failed tests $t1$ and $t2$ are executed by $a$, and $\mathit{np}$ of $f$ includes only one test ($t3$) because it is not run by $f$.

\begin{table}[h]
	\centering
	\caption{Coverage hit spectrum (with four basic statistics) }
	\label{cov_matrix_and_stat}
	\renewcommand{\arraystretch}{1.2}
	\resizebox{0.75\columnwidth}{!}{%
	\begin{tabular}{c|cccc|cccc}
		& t1 & t2 & t3 & t4 & ef  & ep  & nf  & np \\
		\toprule
		a       & 1  & 1  & 1  & 1  & 2   & 2   & 0   & 0  \\
		b       & 1  & 1  & 1  & 1  & 2   & 2   & 0   & 0  \\
		f       & 1  & 0  & 0  & 1  & 1   & 1   & 1   & 1  \\
		g       & 1  & 1  & 1  & 1  & 2   & 2   & 0   & 0  \\
		\bottomrule
		Results & 1  & 1  & 0  & 0  & \multicolumn{4}{l}{}
		\end{tabular}
}
	\vspace*{-5mm}
\end{table}

Most formulae use these four values to determine the location of the bugs as accurately as possible.
In this paper, we use five popular formulae for quantitative evaluation as presented in Table~\ref{SBFLMetrics}. 
The DStar~\cite{metric2}, Ochiai~\cite{metric3} and Tarantula ~\cite{metric4} can be seen as the most popular ones. While the Confidence~\cite{ref4} was used to give importance to suspicious program elements especially for tie-breaking purposes. Finally, the GP13~\cite{metric5} is a \enquote{generated} formula by genetic algorithm which is one of the best performing formulae of this kind.
\begin{table}[H]
	\centering
	\caption{SBFL formulae used in the study}
	\label{SBFLMetrics}
	\resizebox{0.8\columnwidth}{!}{
	\begin{tabular}{c|c}
		Name & Formula  \\
			\toprule
		\conf~(\sconf) & \(\displaystyle \frac{|m_{ef}|}{|m_{ef}|+|m_{nf}|} - \frac{|m_{ep}|}{|m_{ep}|+|m_{np}|}\) \\[0.5cm]	
		\dstar~(\sdstar) &  \(\displaystyle \frac{{|m_{ef}|}^2}{|m_{ep}|+|m_{nf}|} \) \\[0.5cm]	
		\gpA & \(\displaystyle  |m_{ef}| \cdot \left ( 1 + \frac{1}{ 2 \cdot |m_{ep}| + |m_{ef}|} \right ) \) \\[0.5cm]
		\ochiai~(\sochiai) &  \(\displaystyle \frac{|m_{ef}|}{\sqrt{(|m_{ef}|+|m_{nf}|) \cdot (|m_{ef}|+|m_{ep}|)}} \) \\[0.5cm]
		\tarantula~(\starantula) &  \(\displaystyle \frac{\frac{|m_{ef}|}{|m_{ef}|+|m_{nf}|}}{\frac{|m_{ef}|}{|m_{ef}|+|m_{nf}|}+\frac{|m_{ep}|}{|m_{ep}|+|m_{np}|}} \)
	\end{tabular}
	}
		\vspace*{-5mm}
\end{table}

By applying these formulae on the coverage hit spectra of our Java program example in Table~\ref{cov_matrix_and_stat}, we can obtain the suspiciousness scores of each method as presented in Table~\ref{tab:example_scores}. 
It can be noted that in this example, each SBFL formula produces the same suspiciousness score for more than one method. 
In other words, SBFL formulae in this case cannot distinguish the  methods from each other based on their pure scores. 
Hence, the tie problem among program elements affects the SBFL effectiveness in this case.

\begin{table}[H]
	\caption{Program example scores}
	\label{tab:example_scores}
	\centering
	\resizebox{0.75\columnwidth}{!}{%
		\begin{tabular}{c|c|c|c|c|c}
			
			Method & \sconf & \sdstar & \gpA & \sochiai & \starantula \\
			\toprule

			a  & 0.00   & 2.00   & 2.33   & 0.71    & 0.50   \\
			b  & 0.00   & 2.00   & 2.33   & 0.71    & 0.50   \\
			f  & 0.00   & 0.50   & 1.33   & 0.50    & 0.50   \\
			g  & 0.00   & 2.00   & 2.33   & 0.71    & 0.50  

		\end{tabular}
	}

\end{table}

\subsection{Rank Calculations and Ties}

When different elements are assigned the same suspiciousness score, we treat these elements {\em score tied} to each other, and we call any such set of code elements {\em rank ties}.
Clearly, rank ties have at least two elements.
Since the output of SBFL algorithms should be a (weakly) monotone list of ranked code elements according to their suspciousness scores, there are various strategies for dealing with rank ties.
This is especially important when evaluating the effectiveness of an SBFL method in terms of the location of the actually faulty element in the rank list.
In this situation all elements in a rank tie are assigned the same rank value, based on one of these approaches~\cite{ref3}:
\begin{itemize}
	\item \textit{minimum (MIN)}: it refers to the top most position of the elements sharing the same suspiciousness value \textit{(optimistic or the best case)},
	\item \textit{maximum (MAX)}: it refers to the bottom most position \textit{(pessimistic or the worst case)}, or
	\item \textit{average (MID)}: it refers to the medium position of the elements sharing the same suspiciousness value \textit{(average case)}.
\end{itemize}

As a general way of assessing SBFL effectiveness, we will use the average rank approach (Equation~\ref{average_rank_eq}), but we will use the other two options as well to examine tie properties in the sections that follow.
If there are multiple bugs for a program version, the highest rank of faulty elements is used.


\begin{equation}
	\label{average_rank_eq}
	\begin{split}
		\text{MID}  = S + \left ( \frac{\text{E - 1 }}{\text{2 }}\right )
	\end{split}
\end{equation}

Where S is the tie starting position and E is the tie size.\\ 



Table~\ref{tab:avg_ranks_of_example} presents the average ranks of the example program based on several fault localization algorithms.
Ranks that belong to a tie are marked in gray.
It can be stated that two algorithms (\sconf~and \starantula) cannot distinguish the methods from each other at all based on ranks, and the other three approaches result a tie-group that contain 3 methods.

\begin{table}[H]
	\caption{Average ranks of program example}
	\label{tab:avg_ranks_of_example}
	\centering
	\resizebox{0.75\columnwidth}{!}{%
		\begin{tabular}{c|c|c|c|c|c}
			
			Method & \sconf & \sdstar & \gpA & \sochiai & \starantula \\
			\toprule

			a  & \tieGray2.5   & \tieGray2   & \tieGray2   & \tieGray2   & \tieGray2.5   \\
			b  & \tieGray2.5   & \tieGray2   & \tieGray2   & \tieGray2   & \tieGray2.5   \\
			f  & \tieGray2.5   & 4            & 4            & 4            & \tieGray2.5   \\
			g  & \tieGray2.5   & \tieGray2   & \tieGray2   & \tieGray2   & \tieGray2.5  

		\end{tabular}
	}
		\vspace*{-2mm}
\end{table}

Thus, such methods get tied in the ranking and cannot be differentiated from each other in terms of which one has to be examined first.
Therefore, tie-breaking strategies are required to break these ties.
Tie-breaking strategies are not only important to measure the effectiveness of a fault localization formula, they are also important for designing an efficient algorithm.
For example, with an effective strategy, the buggy method can be moved up (i.e. to a better position) in the suspicious list.


%
%


SBFL formulae that do not deal with the issue of rank ties do not take into account other suspiciousness factors derived from the context in the ranking. It is quite frequent that ties include faulty elements and it is not limited to any particular localization technique or target program. As a result, such elements are tied to the same position in the ranking. Also, it gives an indication that the used technique cannot distinguish between the tied elements in terms of their likelihood of being faulty. Thus, no guidance is provided to developers on what to examine first~\cite{ref4}.
In addition, the greater the number of ties involving faulty elements, the more difficult it is to predict at what point the fault will be found during the examination.

Rank ties can be divided into two important types: {\em non-critical} and {\em critical}. 
Non-critical ties refer to the case where only non-faulty elements are tied together for the same score in the ranking. 
Here, if the tied elements have a higher suspiciousness than the actual faulty element, then every element will be examined before finding the fault, regardless of the ties. 
On the other hand, if the tied elements have a lower suspiciousness than the actual faulty element, then the faulty element will be examined before the tied ones. 
Thus, there is no need to continue examining the ranking. 
In either case, the internal order in which the non-critical tied elements are examined does not affect the performance of fault localization in terms of the number of elements that must be examined before finding the fault.

Critical ties, on the other hand, refer to the case when a faulty element is tied with other non-faulty elements~\cite{ref4}. In this type, the internal order of examination affects the performance.
In the case of tie-breaking approaches, critical ties are the main target as it can bring improvement to the efficiency of the SBFL algorithm.
But, unfortunately, we do not know which code element is faulty, so all ties have to be dealt with by the tie-breaking strategy.

\section{Tie Statistics}
\label{tie_breaking}

In this section, we analyze the existence of rank ties in a set of benchmark programs.
We present the subject programs that were used in our experimental results alongside their properties and the granularity of data collection that was employed as  a program  spectra  for our selected SBFL formulae. 
Then, we present the properties of ties we obtained before applying our tie-breaking strategy and to which extent they can be reduced.

\subsection{Subject Programs}

An appropriate dataset is required to examine fault localization. One of the most popular bug dataset is Defects4J.
It is a database of non-trivial real faults which is used to enable reproducible studies in software fault localization for Java programs \cite{ref5}.
Besides, it is the most frequently used benchmark in the fault localization literature \cite {ref9} as it provides a high-level framework interface to easily access faulty and fixed program versions and their corresponding test suites.
The version we used in this study is v1.5.0 and consists of 6 open-source Java programs and 438 real faults which were identified and extracted from the projects’ repositories\footnote{\url{https://github.com/rjust/defects4j/tree/v1.5.0}}.
However, a few faults were excluded in this study due to instrumentation errors or unreliable test results.
Thus, a total of 411 faults were included in the final used dataset. Table \ref{subject_programs} presents each program and its main properties.


\begin{table}[H]
	\centering
	\caption{Subject programs}
	\label{subject_programs}
	\resizebox{0.7\columnwidth}{!}{%
	\begin{tabular}{@{}c|rrrr@{}}
			Project & \begin{tabular}[c]{@{}c@{}}Number\\ of bugs\end{tabular} & \begin{tabular}[c]{@{}c@{}}Size\\ (KLOC)\end{tabular} & \begin{tabular}[c]{@{}c@{}}Number \\ of tests\end{tabular} & \begin{tabular}[c]{@{}c@{}}Number\\ of methods\end{tabular} \\
		\midrule
		Chart   & 25  & 96  & 2.2k  & 5.2k  \\
		Closure & 168 & 91  & 7.9k  & 8.4k  \\
		Lang    & 61  & 22  & 2.3k  & 2.4k  \\
		Math    & 104 & 84  & 4.4k  & 6.4k  \\
		Mockito & 27  & 11  & 1.3k  & 1.4k  \\
		Time    & 26  & 28  & 4.0k  & 3.6k  \\
		\bottomrule
		All   & 411 & 332 & 22.1k & 27.4k  \\
	\end{tabular}
	}
		\vspace*{-4mm}
\end{table}

\subsection{Granularity of Data Collection}
In this paper, method-level granularity was employed as a program spectra or coverage type.
Compared to statement-level granularity, the widely used level, it has several advantages~\cite{method_level}: (a) it provides more comprehensive contextual information about the program entity under investigation, (b) it can handle (i.e., scales well to) large programs and executions, (c) some  studies report  that  it is a better granularity-level for the users too  \cite{empir2, ref8}.
Nevertheless, there is no theoretical obstacle to investigate lower levels of granularity as well in terms of rank ties in the future. 

\subsection{Evaluation Baselines}
In this paper, five standard SBFL formulae, which are presented in Table~\ref{SBFLMetrics}, were used as the baselines to evaluate and compare our proposed method against. The reasons behind this are: (a) there is no other proposed tie-breaking approach that works on the method-level granularity as our method does; (b) our goal was to use contextual information from program executions only to break ties, and not as the underlying SBFL formula for all program elements (this was done by Vancsics et al.~\cite{VHS21}).
The authors in~\cite{ref4} used two confidence formulae to break ties and data-dependency among program statements as well, but these approaches are not directly comparable to ours. 

\subsection{Basic Statistical Analysis}

As mentioned earlier, there is no guarantee that SBFL formulae produce unique suspiciousness scores for all the elements of a program under test. As a result, many elements may share the same scores and get tied with each other. Here, we present brief yet informative statistics on the number of ties that the selected five SBFL formulae produce when applied on the Defects4J dataset (see Table~\ref{tab:num_of_ties}). It can be noted that all the selected SBFL formulae produce ties across all the target programs. This may indicate different things: (a) ties are not rare in fault localization, (b) ties can be formed regardless of which subject program is under consideration, c) different SBFL formulae are affected.

Table~\ref{tab:num_of_crit_ties} presents the number of critical ties. An interesting observation is that the number of ties is not related to program size. For example, smaller programs may have more critical ties than larger programs as in the case of the Lang (22 KLOC) program having more critical ties compared to the Chart (96 KLOC) program. 
The average number of critical ties per bug is an important indicator, as it means in essence the probability that a buggy element will be tied (assuming a single-bug scenario).


\begin{table}[H]
	\caption{Number of ties: total and average per bug}
	\label{tab:num_of_ties}
	\resizebox{1.0\columnwidth}{!}{%
	\begin{tabular}{c|rr|rr|rr|rr|rr}
			\multirow{2}{*}{Project} & \multicolumn{2}{c|}{\sconf} & \multicolumn{2}{c|}{\sdstar} & \multicolumn{2}{c|}{\gpA} & \multicolumn{2}{c|}{\sochiai} & \multicolumn{2}{c}{\starantula} \\
		& \#            & avg            & \#          & avg         & \#         & avg         & \#          & avg          & \#            & avg           \\
		\toprule
		Chart       & 3656     & 146.24   & 508     & 20.32    & 512     & 20.48   & 506     & 20.24    & 490      & 19.60    \\
		Closure     & 86181    & 512.98   & 19069   & 113.51   & 19017   & 113.2   & 19043   & 113.35   & 19109    & 113.74   \\
		Lang        & 3430     & 56.23    & 185     & 3.03     & 188     & 3.08    & 187     & 3.07     & 187      & 3.07     \\
		Math        & 12856    & 123.62   & 844     & 8.12     & 846     & 8.13    & 854     & 8.21     & 851      & 8.18     \\
		Mockito     & 3381     & 125.22   & 779     & 28.85    & 780     & 28.89   & 779     & 28.85    & 789      & 29.22    \\
		Time        & 5776     & 222.15   & 589     & 22.65    & 571     & 21.96   & 597     & 22.96    & 609      & 23.42    \\
		
		\bottomrule
		All         & 115280   & 280.49   & 21974   & 53.46    & 21914   & 53.32   & 21966   & 53.45    & 22035    & 53.61   
	\end{tabular}
}
	\vspace*{-6mm}
\end{table}


\begin{table}[H]
	\caption{Number of critical ties: total and average per bug}
	\label{tab:num_of_crit_ties}
	\resizebox{1.0\columnwidth}{!}{%
		\begin{tabular}{c|rr|rr|rr|rr|rr}
			\multirow{2}{*}{Project} & \multicolumn{2}{c|}{\sconf} & \multicolumn{2}{c|}{\sdstar} & \multicolumn{2}{c|}{\gpA} & \multicolumn{2}{c|}{\sochiai} & \multicolumn{2}{c}{\starantula} \\
			& \#            & avg            & \#          & avg         & \#         & avg         & \#          & avg          & \#            & avg           \\
			\toprule
			Chart   & 14  & 0.56 & 17  & 0.68 & 14  & 0.56 & 15  & 0.6  & 16  & 0.64 \\
			Closure & 102 & 0.61 & 101 & 0.60 & 102 & 0.61 & 101 & 0.60 & 100 & 0.60  \\
			Lang    & 20  & 0.33 & 22  & 0.36 & 20  & 0.33 & 21  & 0.34 & 22  & 0.36 \\
			Math    & 65  & 0.62 & 69  & 0.66 & 65  & 0.62 & 65  & 0.62 & 65  & 0.62 \\
			Mockito & 13  & 0.48 & 13  & 0.48 & 13  & 0.48 & 13  & 0.48 & 13  & 0.48 \\
			Time    & 9   & 0.35 & 9   & 0.35 & 9   & 0.35 & 10  & 0.38 & 10  & 0.38 \\
			
			\bottomrule
			All     & 223 & 0.54 & 231 & 0.56 & 223 & 0.54 & 225 & 0.55 & 226 & 0.55
		\end{tabular}
	}
		\vspace*{-6mm}
\end{table}

We can conclude that more than half of the bugs (54-56\%) are within critical ties, i.e. in most cases there is at least one method whose suspiciousness score is the same as the score of the faulty method.


The sizes of ties is another important factor when considering the potential improvements by tie-breaking.
This can be investigated by looking at the differences between the MIN (best case) and the MID (average case) approaches described in the previous section.
Consider Table~\ref{tab:possible_improvements} which shows the number of critical ties for which MIN and MID values are different in columns 2 and 3 (essentially, the critical tie numbers as shown above),
and also the sum of the corresponding rank differences (column 4), and its average per critical tie (column 5).
Put it differently, the double of the average difference is the average critical tie size in the benchmark, which is around 7 methods.
The difference between the different formulae is not notable.

It also follows that, ideally, the best improvement we could achieve using a tie-breaking technique is these averages.
From Table~\ref{tab:possible_improvements}, we can see in how many cases there is any improvement possible, so we can use these numbers as a baseline for evaluating our tie-breaking approach in subsequent sections.

\begin{table}[H]
	\caption{Improvement possibilities based on critical tie numbers and average tie sizes}
	\label{tab:possible_improvements}
	\centering
	\resizebox{0.8\columnwidth}{!}{
	\begin{tabular}{c|cc|cc}
		& \begin{tabular}[c]{@{}c@{}}MIN != MID\\ (count)\end{tabular} & \begin{tabular}[c]{@{}c@{}}MIN != MID\\ (\%)\end{tabular} & Diff.  & Avg. diff. \\
		\toprule
		\sconf       & 223 & 54.3 & 758.5 & 3.40 \\
		\sdstar      & 231 & 56.2 & 795.0 & 3.44 \\
		\gpA         & 223 & 54.3 & 799.5 & 3.59 \\
		\sochiai     & 225 & 54.7 & 784.0 & 3.48 \\
		\starantula  & 226 & 55.0 & 831.0 & 3.68
	\end{tabular}}
\end{table}

\begin{figure}[H]
	\caption{Distribution of size of critical ties}
	\label{size_distribution}
	\centering
	\includegraphics[width=0.8\columnwidth, height=4cm]{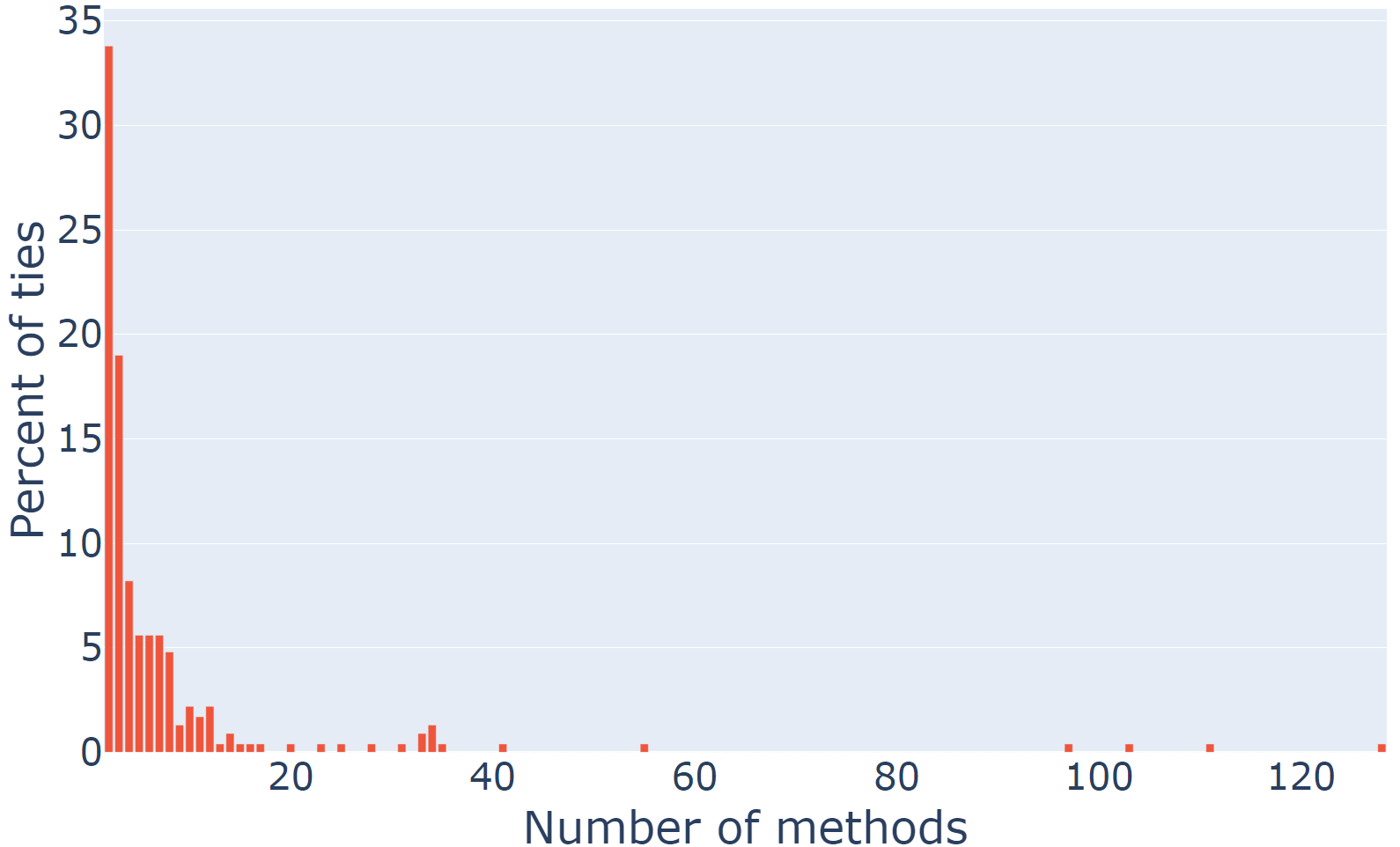}
\end{figure}

We examined the distribution of the critical tie sizes as well, which is shown in Figure~\ref{size_distribution}.
The X-axis represents the number of methods involved in critical ties and the Y-axis represents the percentage of method groups that have the same tie size.
As expected, most ties are relatively small (2--4 elements), 67\% of the critical ties contain 5 or less methods, and sizes above 15 are rare (the average is 7.8, the median 3 and the maximum 128).
Interestingly, there are some outlier cases where the tie sizes are very big, which is the explanation of the relatively large average number.

\noindent\fbox{%
	\parbox{\columnwidth}{%
	\textbf{RQ1}: Overall, it can be said that the ties and critical ties are very common (for the bugs in our benchmark).
	Each of the examined SBFL algorithms created critical ties for more than half of bugs, and on average, the ranks could potentially be improved by around 3.5 positions by eliminating the ties.
	}%
}

\section{Call Frequency-Based Tie-Breaking}
\label{freq_ef}

In this section, we present the concepts of our proposed tie-breaking strategy and how it works. 
Then, we present its effectiveness in reducing critical ties when applied on our bug benchmark.

\subsection{Frequency-based Tie Reduction}
In Section~\ref{fl_ties}, we introduced the basic concept of hit-based SBFL.
One disadvantage of this approach is that it does not take into account the frequency of executing the program elements, in our case methods (also known as the count-based SBFL).
There have been studies that used counts~\cite{harrold1998empirical, harrold2000empirical}, but recent results~\cite{abreu2010exploiting} have shown that these are unable to improve efficiency of the algorithms.

Vancsics et al.~\cite{VHS21} proposed a technique to replace the simple count-based approach that proved to enhance hit-based spectra while eliminating the problems of the naive counts.
It is based on replacing the value of $\mathit{ef}$ in the SBFL formulae with the frequency of different call contexts in the call stack for failing tests.
The basic intuition is that if a method participates in many different calling contexts (both as a caller and as a callee), it will be more suspicious.
In other words, the frequency of methods occurring in the unique call stacks belonging to failing test cases can effectively indicate the location of the bugs.
In the present research, we will employ this concept for the purpose of tie-breaking.
 
To illustrate the basic concept of frequency-based tie reduction, first we define the {\em frequency-based} SBFL matrix that replaces the traditional hit-based one.
In it, each element will get an integer instead of $\{0,1\}$ indicating the number of occurrences of the particular code element in the unique call stacks (effectively, the different contexts) when executing the given test cases~\cite{VHS21}.
Table~\ref{tab:freq-matrix_example} shows the frequency-based matrix for the example.
For instance, the call stacks of $t1$ are $(a,f)$, $(a,g)$ and $(b,g)$, so the frequency of $g$ will be 2 for test $t1$.

\begin{table}[H]
	\caption{Example frequency-matrix}
	\centering
	\label{tab:freq-matrix_example}
	\begin{tabular}{c|cccc|c}
		& a & b & f & g & Results \\
		\toprule
		t1      & 2 & 1 & 1 & 2 & Failed  \\
		t2      & 1 & 1 & 0 & 2 & Failed  \\
		t3      & 1 & 1 & 0 & 1 & Passed  \\
		t4      & 3 & 1 & 1 & 2 & Passed  \\
		\bottomrule
		\sfreqEfNew & 3 & 2 & 1 & 4 &
	\end{tabular}
\end{table}

In the next step, we define our metric to be used as a discriminating factor for tie-breaking. The \sfreqEfNew{} corresponds to the ``frequency-based $\mathit{ef}$'' and is calculated by summarizing the corresponding frequency-based values in the matrix for the failing test cases (see Equation~\ref{freq-ef-eq}).
The values for our example are shown in the last row of Table~\ref{tab:freq-matrix_example}.

\begin{equation}
\label{freq-ef-eq}
	\begin{split}
		\phi(m) & = \sum_{t \in \text{failed test}}c_{m,t}  \\[10pt]
		 m \in \text{methods, } & c_{m,t} \in \text{frequency-matrix} \\[5pt]
	\end{split}
\end{equation}

\begin{figure}[ht]
	\centering
	\includegraphics[width=5cm, height=6.5cm]{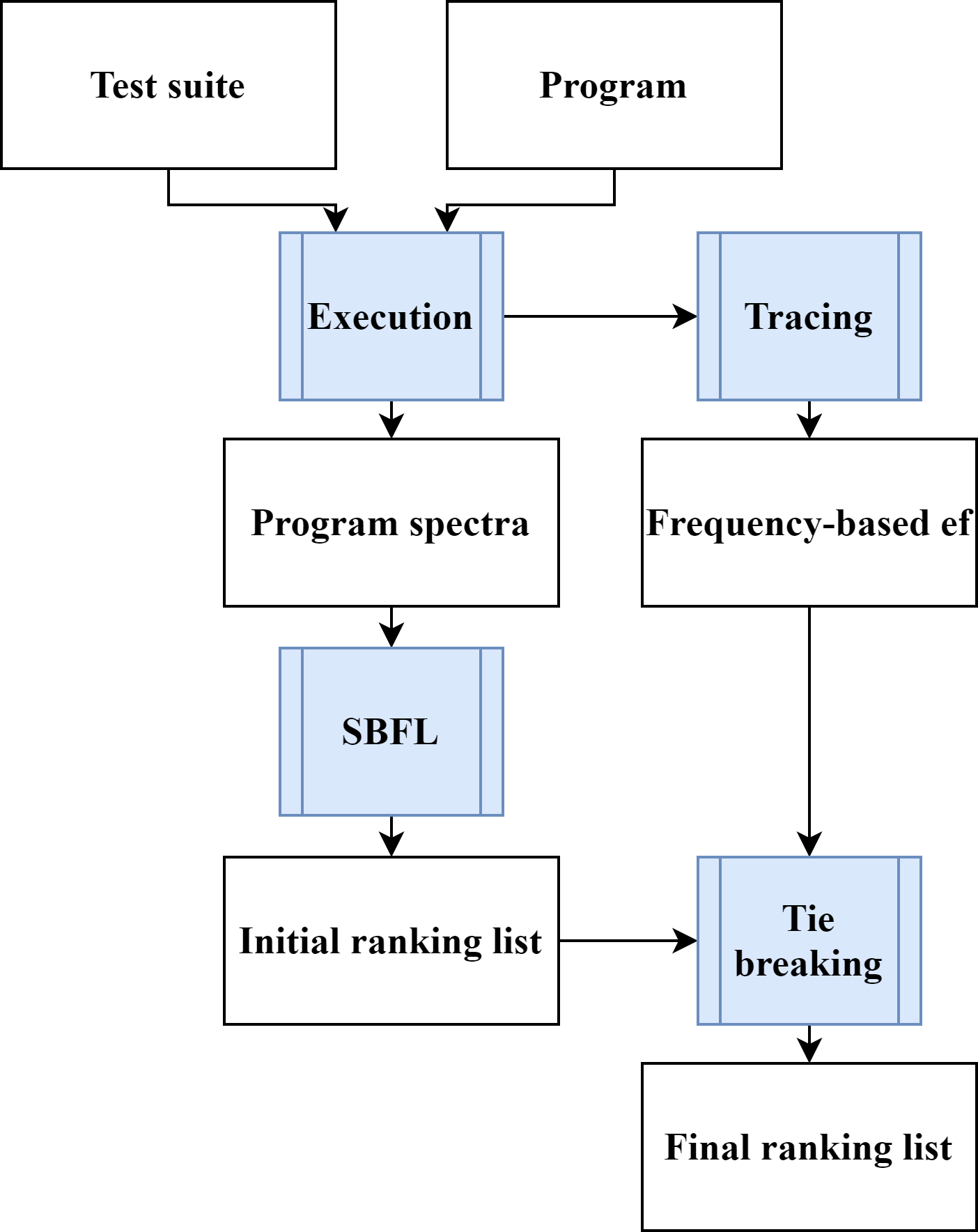}
	\caption{The proposed tie-breaking process}
	\label{TieBreaking_Process}
	\vspace*{-5mm}
\end{figure}

Figure~\ref{TieBreaking_Process} shows our tie-breaking process which can be seen as a two-stage process. In the first stage, we compute the suspiciousness scores of program methods and their ranks via applying different SBFL formulae on the program spectra (test coverage and test results). The output of this stage is an initial ranking list of program methods including critical and non-critical ties. In the second stage, we trace the execution of program methods to obtain the \sfreqEfNew{}, i.e. frequency-based $\mathit{ef}$. This will then be used as a tie-breaker after re-arranging the order of the critical tied methods in the initial ranking list based on the value of \sfreqEfNew{} for each method. The output of this stage is a final ranking list, where many critical ties either eliminated completely or their sizes were reduced.  

Our proposed tie-breaking method uses the obtained \sfreqEfNew{} call frequency values to break the methods sharing the same score, by putting the methods with higher \sfreqEfNew{} upper in the rank. Thus, the most suspicious one will be the method that was called in more different call stacks from failing test cases. The rationale behind using the \sfreqEfNew{} rather than other contexts (such as the context of method calls in passing test cases) is the intuition that a method is more suspicious to contain a fault when executed by more failing test cases than passing ones, while non-suspicious when mostly executed  by passing tests. However, other different contexts could be considered in the future and investigate their impacts on breaking ties.

The ranks without (columns B) and with tie-breaking (columns A) with our approach for the example are presented in Table~\ref{tab:rank_example}.
Ties marked in gray were eliminated with the use of call frequency. As a result, we were able to differentiate between the faulty method ($g$) and the other suspicious ones using all of the SBFL formulae (the faulty method got the highest rank in all cases).
\begin{table}[H]
	\caption{Ranks \textbf{B}efore and \textbf{A}fter using the tie-breaking strategy}
	\label{tab:rank_example}
	\centering
	\resizebox{0.9\columnwidth}{!}{%
		\begin{tabular}{c|c|cc|cc|cc|cc|cc}
				\multirow{2}{*}{Method} & \multirow{2}{*}{\sfreqEfNew{}} & \multicolumn{2}{c|}{\sconf} & \multicolumn{2}{c|}{\sdstar} & \multicolumn{2}{c|}{\gpA} & \multicolumn{2}{c|}{\sochiai} & \multicolumn{2}{c}{\starantula} \\[0.4cm]
			&                          & B            & A         & B           & A           & B           & A          & B            & A           & B           & A         \\
			\toprule
			a   & 3   & \tieGray2.5     & 2         & \tieGray2           & 2           & \tieGray2           & 2          & \tieGray2            & 2           & \tieGray2.5         & 2         \\
			b   & 2   & \tieGray2.5     & 3         & \tieGray2           & 3           & \tieGray2           & 3          & \tieGray2            & 3           & \tieGray2.5         & 3         \\
			f   & 1   & \tieGray2.5     & 4         &         4           & 4           &         4           & 4          &         4            & 4           & \tieGray2.5         & 4         \\
			g   & 4   & \tieGray2.5     & 1         & \tieGray2           & 1           & \tieGray2           & 1          & \tieGray2            & 1           & \tieGray2.5         & 1        
		\end{tabular}
	}
\end{table}

\subsection{Reduction of the Critical Ties}

The metric called \textit{Tie-reduction} was defined by Xu et al. to measure how much a critical tie has been reduced/broken in terms of size~\cite{ref4}.
Here, size simply means the number of code elements sharing the same score value, and obviously, the minimum tie size is 2.
The goal of any tie-breaking strategy is to reduce the size of the tie or completely eliminate it (when the resulting size is 1).
We modified the original definition of this metric to better reflect the actual gain in terms of what portion of the ``superfluous'' elements in a tie can be eliminated (see Equation~\ref{tie_red_eq}).

\begin{equation}
	\label{tie_red_eq}
	\begin{split}
		\text{Tie-Reduction} & = \left ( 1 - \frac{\text{size}_{\text{after}}-1}{\text{size}_{\text{before}}-1}\right ) \cdot 100\%
	\end{split}
\end{equation}

\noindent Here, size$_\text{after}$ is the size of a critical tie after applying a tie-breaking strategy and size$_\text{before}$ is the size of a critical tie before applying a tie-breaking strategy.
In an ideal case, the critical tie is completely eliminated, in which case the value of the tie-reduction is $100\%$.
If no reduction can be obtained, this metric will be $0\%$, and in all other cases it will show the percentage of the removed elements that share the same score value as the faulty element.

\begin{figure}[H]
	\caption{Tie-reduction distribution of critical ties}
	\label{tie_red_figure}
	\centering
	\includegraphics[width=0.8\columnwidth, height=4cm]{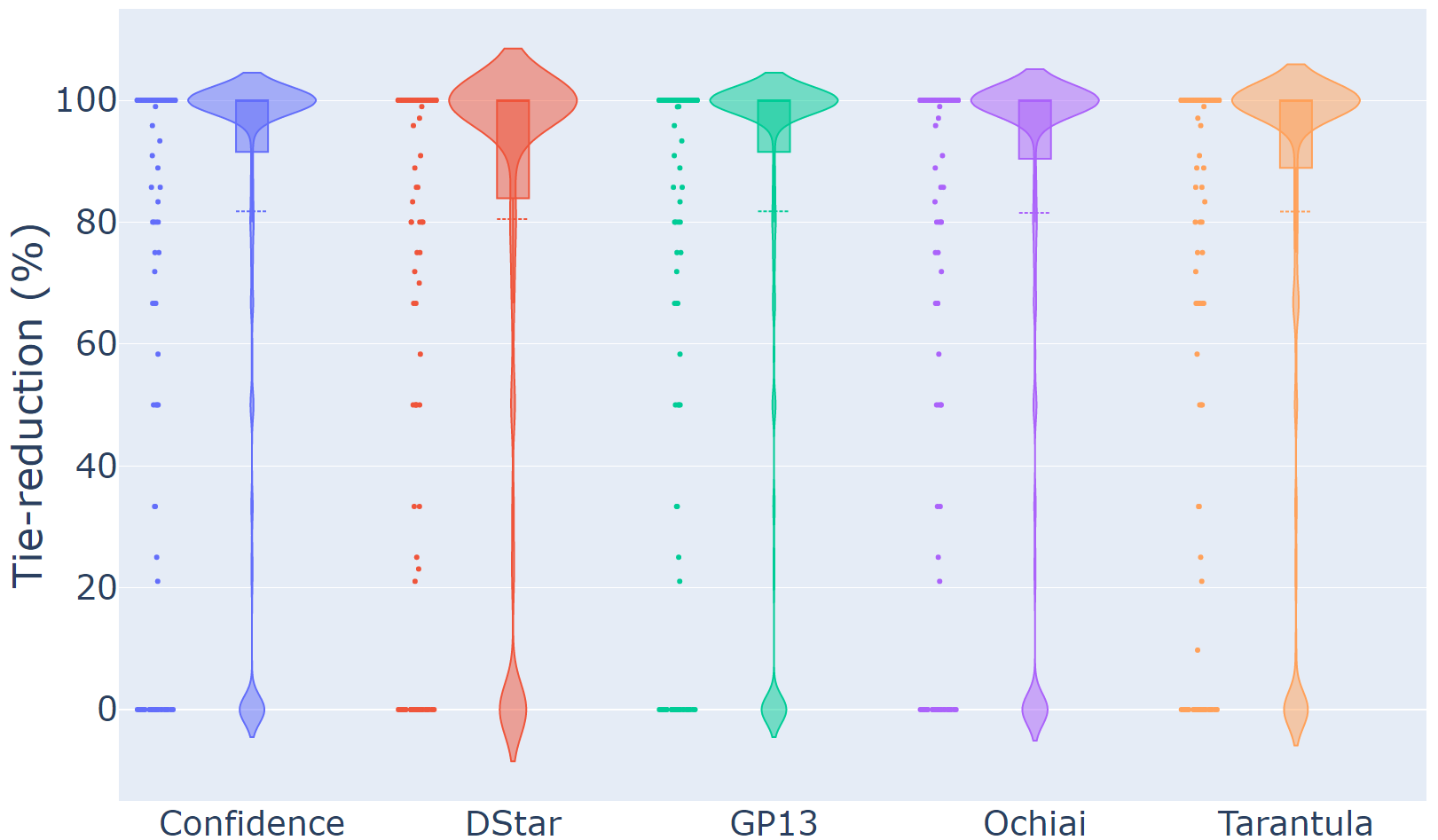}
	\vspace*{-3mm}
\end{figure}

In Figure~\ref{tie_red_figure}, we visualized the amount of critical tie-reduction on our benchmark using the Tie-Reduction metric.
Each dot represents one bug in the dataset and the violin plot offers a more general picture about the distribution of the data points.
It can be seen from the shape of the plots that in several cases, reduction was not possible but the majority of the ties was completely eliminated.
Similar to the number and size of critical ties, there was no significant difference in this aspect depending on what SBFL formula was used, we obtained very similar results.

Table~\ref{tab:tie_red_stats} shows some important statistical values for this dataset: mean, median and quartile 1 (the value in the middle between the smallest and the median points).
Since the median is 100\%, we can state that the critical ties are eliminated by our method in more than half of the cases (72--73\%, as detailed below), and the reduction is between 83.9--91.5\% for three-quarters of the bugs, while the average rate of reduction is greater than 80\% in all cases.

\begin{table}[H]
	\caption{Statistics of tie-reduction (in percentage)}
	\label{tab:tie_red_stats}
	\centering
	\resizebox{0.8\columnwidth}{!}{%
	\begin{tabular}{c|rrrrrr}
		              & \sconf & \sdstar & \gpA  & \sochiai & \starantula   \\
		              \toprule
		Mean          & 81.8   & 80.5    & 81.8  & 81.5     & 81.8  \\
		Median        & 100.0  & 100.0   & 100.0 & 100.0    & 100.0 \\
		$\mathit{Q1}$ & 91.5   & 83.9    & 91.5  & 90.4     & 88.9  \\
	\end{tabular}
	}
\end{table}

Table~\ref{tab:crit_tie_diff} presents the number of remaining critical ties for each program and SBLF formula after applying the tie-reduction algorithm.
The difference to the previous values (shown in Table~\ref{tab:num_of_crit_ties}) is also included.
For example, 15 bugs of Chart were in critical ties with the \ochiai~formula, but after applying the call frequency-based tie-reduction 11 critical ties are eliminated, which is 73.3\% of the initial ties.
Overall, we achieved 72-73\% improvement in the number of critical ties over the full dataset, the best case being Mockito with over 84.6\% and the worst result was 54.5\% on Lang using \tarantula~and \dstar.

\begin{table}[H]
	\caption{Changes in the number of critical ties after reduction}
	\label{tab:crit_tie_diff}
	\centering
	\resizebox{0.9\columnwidth}{!}{%
		\begin{tabular}{cc|cccccc|c}
			&           & Chart & Closure & Lang & Math & Mockito & Time & All \\
			\toprule
			\multirow{3}{*}{\sconf} & after     & 4 & 24 & 9 & 18 & 2 & 2  & 59 \\
			& diff. (\#) & 10 & 78 & 11 & 47 & 11 & 7  & 164 \\
			& diff. (\%) & 71.4 & 76.5 & 55.0 & 72.3 & 84.6 & 77.8  & 73.5 \\
			\midrule	
			\multirow{3}{*}{\sdstar}  & after     & 6 & 24 & 10 & 21 & 2 & 2  & 65 \\
			& diff. (\#) & 11 & 77 & 12 & 48 & 11 & 7  & 166 \\
			& diff. (\%) & 64.7 & 76.2 & 54.5 & 69.6 & 84.6 & 77.8  & 71.9 \\
			\midrule
			\multirow{3}{*}{\gpA} & after     & 4 & 25 & 9 & 18 & 2 & 2  & 60 \\
			& diff. (\#) & 10 & 77 & 11 & 47 & 11 & 7  & 163 \\
			& diff. (\%) & 71.4 & 75.5 & 55.0 & 72.3 & 84.6 & 77.8  & 73.1 \\
			\midrule
			\multirow{3}{*}{\sochiai}  & after     & 4 & 24 & 9 & 19 & 2 & 2  & 60 \\
			& diff. (\#) & 11 & 77 & 12 & 46 & 11 & 8  & 165 \\
			& diff. (\%) & 73.3  & 76.2 & 57.1 & 70.8 & 84.6 & 80.0 & 73.3 \\
			\midrule
			\multirow{3}{*}{\starantula}  & after     & 5 & 24 & 10 & 19 & 2 & 2  & 62 \\
			& diff. (\#) & 11 & 76 & 12 & 46 & 11 & 8  & 164 \\
			& diff. (\%) & 68.8 & 76.0 & 54.5 & 70.8 & 84.6 & 80.0  & 72.6 \\           
		\end{tabular}
	}
\end{table}


The sizes of the critical ties determine the level of achievable improvement after applying a tie-breaking approach. 
However, it is also important in which direction in the new ranking the faulty element moved after tie-breaking.
Using the terminology from the previous section, moving from the MID position towards MIN means improvement.
In the previous section, in Table~\ref{tab:possible_improvements}; we presented the maximum potential improvement that sets a theoretical constraint on SBFL effectiveness after tie-reduction. Table~\ref{tab:min_achieving} presents what we actually achieved using our proposed algorithm (the meaning of the data is the same as in Table~\ref{tab:possible_improvements}).

\begin{table}[H]
	\caption{Achieving the minimum ranks}
	\label{tab:min_achieving}
	\centering
	\resizebox{0.8\columnwidth}{!}{
		\begin{tabular}{c|cc|cc}
			& \begin{tabular}[c]{@{}c@{}}MIN != MID\\ (count)\end{tabular} & \begin{tabular}[c]{@{}c@{}}MIN != MID\\ (\%)\end{tabular} & Diff.  & Avg. diff. \\
			\toprule
			\sconf 		& 59 & 14.4 & 53.5 & 0.9 \\
			\sdstar 	& 65 & 15.8 & 61.5 & 0.9 \\
			\gpA 		& 60 & 14.6 & 54.0 & 0.9 \\
			\sochiai	& 60 & 14.6 & 55.5 & 0.9 \\
			\starantula	& 62 & 15.1 & 94.0 & 1.5
		\end{tabular}
	}
\end{table}

We examined whether our method was able to reduce the number of cases where the MIN (best case) and the MID (average case) approaches give different results.
If there were no such cases that would mean that the obtained new ranking after tie-break would always be the best possible, MIN case.
Table~\ref{tab:min_achieving} shows that, after our approach, only around 15\% of the bugs contained critical ties (column 3), compared to around 55\% before tie-breaking.

Comparing this with the result of Table~\ref{tab:possible_improvements}, we find that in more than 160 cases we managed to achieve the ideal result with our method where the original algorithm was not able to do so.
It means that for nearly three quarters of bugs in critical ties (72--73\%), the non-optimal result was improved to optimal.

If we compare the sum of the rank differences (column 4) and their averages (column 5) in Tables~\ref{tab:possible_improvements} and~\ref{tab:min_achieving}, it can be seen that our approach was able to reduce the sum significantly (by 89--93\%), and the average by 59--74\%.
Put it differently, the overall rank positions from the ideal case improved from around 3.5 to 1 in the cases when we achieved optimal result, which essentially means rank improvement between $2.2$ and $2.5$.\\

\noindent\fbox{%
	\parbox{\columnwidth}{%
		\textbf{RQ2}: Using the call-frequency based tie-breaking strategy, we achieved a significant reduction in both size and number of critical ties in our benchmark. In 72-73\% of the cases the ties were completely eliminated, the average reduction rate being more than 80\%. In nearly three quarters of the cases (72--73\%), the faulty element got the highest rank among the tie-broken code elements, and here it improved its position by 59--74\% on average.
	}%
}

\section{Effect of Tie-Breaking on SBFL Performance}
\label{results}

In this section, we analyze what is the overall effect of the proposed tie-breaking strategy on SBFL effectiveness in terms of global ranks.
For that purpose we use several evaluation metrics that were employed in the literature~\cite{ref10, ref11, ref4}.

\subsection{Achieved improvements and the average ranks}

Average rank is used to rank the program elements that share the  same  suspiciousness  value by considering the average of their positions after they get sorted, in a descending order, by the level of their suspiciousness. And, it  is  calculated  using Equation 1.
Table~\ref{tab:avg_ranks} presents the average ranks before (column 2) and after (column 3) applying our tie-breaking strategy and it shows the difference between the average ranks before and after tie-reduction (column 4). 
If the difference is negative then this means that we could achieve improvement with our proposed strategy. 

\begin{table}[H]
	\caption{Average rank of faulty elements before and after tie-breaking}
	\label{tab:avg_ranks}
	\centering
	\resizebox{0.5\columnwidth}{!}{%
	\begin{tabular}{c|ccc}
		            & Before & After & Diff. \\
		            \toprule
		\sconf      & 55.16  & 53.11 & -2.05 \\
		\sdstar     & 46.86  & 44.79 & -2.07 \\
		\gpA        & 68.79  & 66.68 & -2.11 \\
		\sochiai    & 46.95  & 44.81 & -2.14 \\
		\starantula & 50.39  & 48.33 & -2.06
	\end{tabular}
	}
		\vspace*{-3mm}
\end{table}

We can see that our strategy achieved improvements with all the selected SBFL formulae: the average rank reduced by more than $2$ in all cases, which corresponds to 3.1--4.1\% with respect to the total number of elements. 
Note, that this average is similar to what we got for RQ2, but it is not the same because for RQ2 we examined only the cases when we achieved the optimal result, while in this section we are interested in the global results.

We also examined how many times our tie-breaking strategy changed the rank of bugs (in positive and negative directions) and what was the impact of the changes.
Table~\ref{avg_ranks_compare} presents the possible changes in several categories, as follows (B means before, A means after applying tie-breaking):
\begin{enumerate}
	\item the faulty method moved to the top of the critical tie (column: best), when
	$\text{B}^{\text{MIN}} = \text{A}^{\text{MID}}$ (this is the case that we discussed using Tables~\ref{tab:possible_improvements} and~\ref{tab:min_achieving})
	\item it has moved up in the rankings (column: better), when
	$\text{B}^{\text{MID}} > \text{A}^{\text{MID}}$ and $\text{B}^{\text{MIN}} < \text{A}^{\text{MID}}$
	\item it remained in the same position (column: same), when
	$\text{B}^{\text{MID}} = \text{A}^{\text{MID}}$
	\item we worsened the result (column: worse), when\\
	$\text{B}^{\text{MID}} < \text{A}^{\text{MID}}$ and $\text{B}^{\text{MAX}} > \text{A}^{\text{MID}}$
	\item it slipped back to the worst place (column: worst), when
	$\text{B}^{\text{MAX}} = \text{A}^{\text{MID}}$
\end{enumerate}

\noindent In addition, column ``improve'' represents improvements in rank modifications (i.e., best+better), while ``deteriorate'' is worse+worst.
The table also includes the average differences in rank positions for the given categories.

The results indicate that in about 3--4 times more cases we achieved improvement than deterioration of the ranking results.
Moreover, the improvement differences are much higher than the deterioration differences (compare, for example, the {\em better} cases of around -7 to {\em worse} cases of around 2).
Other interesting insight is that in the case of \textit{best}, the difference is relatively small as the size of the ties broken in this category were small as well (they contained 3-4 methods).
Looking at the overall result, the average rate of improvement ranged from -3.73 to -3.86, while the deterioration was only between 1.34 and 1.54 rank positions on average.

\begin{table}[H]
	\caption{Comparison of average ranks before and after tie-breaking}
	\label{avg_ranks_compare}
	\centering
	\resizebox{\columnwidth}{!}{
	\begin{tabular}{cc|rr|r|rr|rr}
			\multicolumn{2}{l|}{} & Best       & Better     & Same     & Worse     & Worst     & Improve     & Disimprove  \\
		\toprule
		\multirow{2}{*}{\sconf} & count     & 85    & 51    & 50 & 17   & 20   & 136   & 37   \\
		& avg. diff. & -1.71 & -7.13 & 0  & 2.32 & 0.55 & -3.74 & 1.36 \\
		\midrule
		\multirow{2}{*}{\sdstar}      & count     & 85    & 53    & 53 & 18   & 22   & 138   & 40   \\
		& avg. diff. & -1.71 & -7.32 & 0  & 2.31 & 0.55 & -3.86 & 1.34 \\
		\midrule
		\multirow{2}{*}{\gpA}       & count     & 85    & 51    & 50 & 17   & 20   & 136   & 37   \\
		& avg. diff. & -1.71 & -7.39 & 0  & 2.32 & 0.55 & -3.84 & 1.36 \\
		\midrule
		\multirow{2}{*}{\sochiai}     & count     & 86    & 52    & 50 & 16   & 21   & 138   & 37   \\
		& avg. diff. & -1.70  & -7.42 & 0  & 2.38 & 0.62 & -3.86 & 1.38 \\
		\midrule
		\multirow{2}{*}{\starantula}  & count     & 83    & 58    & 47 & 17   & 21   & 141   & 38   \\
		& avg. diff. & -1.46 & -6.97 & 0  & 2.68 & 0.62 & -3.73 & 1.54 \\
	\end{tabular}
}
	\vspace*{-3mm}
\end{table}

The overall rank position improvement might seem modest, but we must consider the fact that the improvement can be achieved only by rearranging the positions in the critical ties. Thus, the sizes of the critical ties serve as a hard constraint (as discussed in the previous section).
However, there is a class of improvements which are probably more important than the general case: improvements in the \textit{Top-N} rank positions, and here the benefits are more pronounced, as presented below.

\subsection{Top-N categories}

Several studies \cite {ref12, ref13} report that developers think that inspecting the first 5 program elements in the ranks list produced by a fault localization technique is acceptable and that the first 10 elements are the upper bound for inspection before ignoring the ranks list.
Hence, we verified the results by focusing on these rank positions only (collectivelly called \textit{Top-N}).
We will use five cases: where a fault is ranked first (Top-1), it is less or equal to three (Top-3), less or equal to five (Top-5), less or equal to ten (Top-10), and when it is more than ten (Other).

We also used a special non-accumulating variant of Top-N categories, in which case we counted cases where the bug fell into a non-overlapping intervals of $[1]$, $(1, 3]$, $(3, 5]$, $(5, 10]$ or $(10, . . .]$.
The goal of the evaluation in this part was to see in how many cases our approach \enquote{moves} a bug into a better (for example, from $(5,10]$ to $(1,3]$) or a worse (for example, from $[1]$ to $(1,3]$) group.
In other words, in how many cases do the bugs get into a higher-rank group (this kind of improvement is also known as {\em enabling improvement}~\cite{ref11}) and in how many cases do we downgrade the category.


Table  \ref{tab:topN}  presents the number of bugs belonging to the corresponding Top-N categories (cumulative) with their percentages, 
for the whole dataset, before and after applying our tie-breaking strategy, as well as the differences between them.
A decrease in the number of cases of the Other category and increase in any Top-N means improvement.

\begin{table}[H]
	\caption{Top-N categories}
	\label{tab:topN}
	\centering
	\resizebox{1.0\columnwidth}{!}{
		\begin{tabular}{c|rr|rr|rr|rr|rr}
			\multirow{2}{*}{} & \multicolumn{2}{c|}{Top-1}                               & \multicolumn{2}{c|}{Top-3} & \multicolumn{2}{c|}{Top-5} & \multicolumn{2}{c|}{Top-10} & \multicolumn{2}{c}{Other} \\
			& \# & \% & \#      & \%      & \#      & \%      & \#       & \%      & \#      & \%      \\
			\toprule
			\sconf              & 75       & 18.2     & 169      & 41.1     & 203      & 49.4     & 246       & 59.9      & 165       & 40.1      \\
			After tie-breaking & 92       & 22.4     & 180      & 43.8     & 214      & 52.1     & 252       & 61.3      & 159       & 38.7      \\
			Diff.               & 17       & 22.7     & 11       & 6.5      & 11       & 5.4      & 6         & 2.4       & -6        & -3.6      \\
			\bottomrule
			
			\sdstar             & 65       & 15.8     & 172      & 41.8     & 210      & 51.1     & 249       & 60.6      & 162       & 39.4      \\
			After tie-breaking &  84    &	20.5    &	186     &	45.4    &	222     &	54.1    &	257   &	62.7      &	153       &	37.3      \\
			Diff.               & 19       & 29.2     & 14       & 8.1      & 12       & 5.7      & 8         & 3.2       & -8        & -4.9      \\
			\bottomrule
			
			\gpA               & 75       & 18.2     & 169      & 41.1     & 203      & 49.4     & 245       & 59.6      & 166       & 40.4      \\
			After tie-breaking & 92       & 22.4     & 179      & 43.6     & 212      & 51.6     & 250       & 60.8      & 161       & 39.2      \\
			Diff.               & 17       & 22.7     & 10       & 5.9      & 9        & 4.4      & 5         & 2.0       & -5        & -3.0      \\
			\bottomrule
			
			\sochiai            & 68       & 16.5     & 173      & 42.1     & 210      & 51.1     & 250       & 60.8      & 161       & 39.2      \\
			After tie-breaking & 87       & 21.2     & 186      & 45.3     & 222      & 54.0     & 257       & 62.5      & 154       & 37.5      \\
			Diff.               & 19       & 27.9     & 13       & 7.5      & 12       & 5.7      & 7         & 2.8       & -7        & -4.3      \\
			\bottomrule
			
			\starantula         & 65       & 15.8     & 166      & 40.4     & 203      & 49.4     & 244       & 59.4      & 167       & 40.6      \\
			After tie-breaking & 83       & 20.2     & 177      & 43.1     & 212      & 51.6     & 251       & 61.1      & 160       & 38.9      \\
			Diff.               & 18       & 27.7     & 11       & 6.6      & 9        & 4.4      & 7         & 2.9       & -7        & -4.2

	\end{tabular}}
		\vspace*{-4mm}
\end{table}

It can be clearly seen that our proposed tie-breaking strategy achieves improvements in all categories by moving many bugs to higher ranked categories.
On the lower end of the scale (Other category with rank $>10$), 5--8 bugs were moved into one of the Top-N categories.
This is important as it brings a “new hope” that a bug could be found by the user with the proposed strategy while it was not very probable without it.
We can see a quite large number of improvements in higher categories as well, around 18 bugs moved to Top-1, for instance.
Note that the percentages of bugs in each category before and after applying the strategy were calculated with respect to the total number of bugs in the dataset. While
the difference percentage was calculated with respect to the number of bugs before applying the strategy.



To better understand the actual changes between the different Top-N categories we should use the non-accumulating variant of these categories.
This shows whether there has been a beneficial change in the rank category.
These moves between the Top-N categories are presented by Table~\ref{tab:topN_move_count}.
The sign \xmark~indicates the number of changes in the negative direction (worsening result), and \cmark~marks improvement.
For example, there were a total of 2 bugs with a rank greater than 1 but less than or equal to 3 before reduction by \tarantula, but our method resulted in a rank value greater than 3 (this is a negative result).
In contrast, our method gave a rank of 1 for the faulty method 15 times which was previously greater than 1 but smaller than 3 (using \tarantula).

These numbers clearly show that improvement was dominant: degradation by the proposed method was observable only for 2--3 bugs in the dataset, while we observed positive changes for 36--44 bugs.


\begin{table}[H]
	\caption{Top-N moves}
	\label{tab:topN_move_count}
	\centering
	\resizebox{0.9\columnwidth}{!}{%
	\begin{tabular}{c|llllllll|ll}

			& $[1]$ & $(1,3]$  & $(1,3]$  & $(3,5]$  & $(3,5]$  & $(5,10]$  & $(5,10]$  & $Other$  & \xmark & \cmark 
			\\ & $\downarrow$ & $\downarrow$  & $\downarrow$  & $\downarrow$  & $\downarrow$  & $\downarrow$  & $\downarrow$  & $\downarrow$  &  & 
			\\ & \xmark & \xmark  & \cmark  &  \xmark &  \cmark & \xmark  &  \cmark & \cmark  &  & \\
		\toprule
		\sconf       & 0   & 1    & 13    & 1    & ~8     & 0     & 11    & 6     & 2     &    38 \\
		\sdstar      & 0   & 1    & 15    & 1    & 10     & 0     & 11    & 8     & 2     &    44 \\
		\gpA         & 0   & 1    & 13    & 1    & ~8     & 0     & 10    & 5     & 2     &    36 \\
		\sochiai     & 0   & 1    & 15    & 1    & ~9     & 1     & 11    & 8     & 3     &    43 \\
		\starantula  & 0   & 2    & 15    & 1    & 11     & 0     & ~8    & 7     & 3     &    41
	\end{tabular}
	}
\end{table}

\noindent\fbox{%
	\parbox{\columnwidth}{%
		\textbf{RQ3}: The efficiency of all investigated SBFL formulae could be improved by using the proposed tie-breaking strategy: the average improvement of rank values in the benchmark was about {\em two positions}, and about 3-4 times more frequenty we observed improvement than detoriation, such improvements being much higher as well.
		Considering the Top-N categories, notable improvements could be observed:
		all Top-N categories showed positive results (improvements in 36--44 cases), 
		and at the same time, in only a few (2--3) cases Top-N categories worsened.
		We were able to increase the number of cases where the faulty method became the top ranked element by 23--30\%. 
	}%
}

\section{Threats to validity}
\label{threats}

Various threats may affect the validity of experimental studies in software engineering. In our work, we considered the following actions to avoid or minimize such threats:

\begin{itemize}

\item Selection of evaluation metrics:	
to ensure that our results and corresponding conclusions are valid, we selected several evaluation metrics that are also used by previous research to ensure multiple-dimension comparisons. 
Besides, all the evaluation metrics employed in this study were reported and described in detail.

\item Correctness of implementation:
to ensure that our experiment implementation is correct and accurate, code review was conducted by the authors. Furthermore, we have run our proposed approach several times to ensure that it is implemented correctly. 

\item Selection of subject programs:
in our experiment, we evaluated the effectiveness of the proposed tie-breaking strategy on fault localization using only six Java subject programs. Thus, we cannot generalize our findings to other programs in general. However, we believe that the selected subject programs are representative to others as they have real faults, varying in size and complexity, and the benchmark containing them, Defects4J, is used commonly in other studies on software fault localization. 

\item Exclusion of faults:
in our experiment, 27 faults (about 6\% of the total faults) of the Defects4J dataset were excluded because we could not compute their call stack information due to technical limitations. The issue here is whether other researchers using the same dataset will be able to replicate our findings. This exclusion was in no ways influenced by the results of the used metrics and the excluded faults are distributed in the dataset approximately evenly, so we believe that this risk can be considered minimal.

\item Selection of SBFL formulae:
to evaluate the effectiveness of our proposed tie-breaking strategy on fault localization, we selected a set of five SBFL formulae in our experiment, which is just a fraction of the proposed techniques in literature. 
The obtained results show improvements with all of them. 
However, we cannot guarantee that the same improvements can be obtained by using other SBFL formulae. 
To mitigate the effect of this issue, we selected three SBFL formulae that are commonly used in other studies on software fault localization, which we extended with two special kinds, one of which was especially designed with tie-breaking in mind.

\end{itemize}

\section{Conclusion}
\label{conclusions}

Rank ties in SBFL are very common regardless of the formula employed, and by breaking these ties, improvements to the localization effectiveness can be expected.
This paper proposes the use of method call contexts for breaking critical ties in SBFL. We rely on instances of call stack traces, which are useful software artifacts during run-time and can often help developers in debugging.
The frequency of the occurrence of methods in different call stack instances determines the position of the code elements within the set of other methods tied together by the same suspiciousness score.

Experimental results show that the proposed tie-breaking strategy, using the Defects4J benchmark, (a) completely eliminated many critical ties with significant reduction of others, and (b) achieved improvements in average rank positions for all investigated SBFL formulae with moving many bugs to the highest Top-N rank positions.
However, there are limits to how much improvement one can expect from tie-breaking alone (we analyzed this limit in the paper and compared to the results achieved).
This means that no matter how clever a tie-breaking method is, it cannot rearrange code elements outside of the tied ranking positions.
Since ties seem to be prevalent, it could be an interesting further research to devise specific tie-aware approaches or modified formulae that minimize ties in the scores and/or break them automatically.

As other future work, we would like to measure the effectiveness of the proposed tie-breaking strategy on other levels of granularity such as statement, branch, etc. 
Employing other SBFL formulae across a much broader range of programs in terms of numbers, types, sizes, and used programming languages, to capture the ties problem characteristics and identify what factors affect them would be interesting for further investigation. 
We also would like to tackle the ties problem by employing other contextual factors beyond method calls and to measure their impacts on the SBFL.  

The results of our experimental study can be found at "https://bit.ly/3qFQmof".

\ifanonym


\else

\section*{Acknowledgements}
The research was supported by the Ministry of Innovation and Technology, NRDI Office, Hungary within the framework of the Artificial Intelligence National Laboratory Program, and by grant NKFIH-1279-2/2020 of the Ministry for Innovation and Technology. Qusay Idrees Sarhan was supported by the Stipendium Hungaricum scholarship programme.

\fi

\balance

\bibliographystyle{IEEEtran}
\bibliography{IEEEabrv,references}

\end{document}